\documentclass{PoS}

\let\OLDthebibliography\thebibliography
\renewcommand\thebibliography[1]{
  \OLDthebibliography{#1}
  \setlength{\parskip}{0pt}
  \setlength{\itemsep}{0pt plus 0.0ex}
}

\makeatletter
\newcommand{\figcaption}[1]{\def\@captype{figure}\caption{#1}}
\newcommand{\tblcaption}[1]{\def\@captype{table}\caption{#1}}
\makeatother

\title{{\vspace{-38mm} \normalsize\hfill{\small DESY 14-230}}\\[38mm]
{\vspace{-41mm}\normalsize\hfill{\small SFB/CPP-14-96}}\\[10mm]The temperature dependence of the chiral condensate in the Schwinger model with Matrix Product States}
\ShortTitle{The chiral condensate in the Schwinger model with Matrix Product State}
\author{\speaker{H.~Saito}$^a$, M.~C.~Ba{\~n}uls$^b$, K.~Cichy$^{acd}$,  J.~I.~Cirac$^b$, K.~Jansen$^a$\\
    \llap{$^a$}John von Neumann Institute for Computing (NIC), DESY, Platanenallee 6, \\
    15738 Zeuthen, Germany\\
    \llap{$^b$}Max Planck Institute of Quantum Optics, Hans Kopfermann 1, 85748, Garching, Germany\\
    \llap{$^c$}Goethe-Universit{\"a}t Frankfurt am Main, Institut f{\"u}r Theoretische Physik, \\
    Max-von-Laue-Stra{\ss}e 1, D-60438 Frankfurt am Main, Germany\\
    \llap{$^d$}Faculty of Physics, Adam Mickiewicz University, Umultowska 85, 61-614 Pozna\'{n}, Poland\\
    E-mail: \email{hana.saito@desy.de}
}
\abstract{
We present our recent results for the tensor network (TN) approach to lattice gauge
theories. 
TN methods provide an efficient approximation for quantum many-body states. 
We employ TN for one dimensional systems, Matrix Product States, to investigate the 1-flavour Schwinger
model. 
In this study, we compute the chiral condensate at finite temperature. 
From the continuum extrapolation, we obtain the chiral condensate in the high temperature region 
consistent with the analytical calculation by Sachs and Wipf. 
}
\FullConference{The 32nd International Symposium on Lattice Field Theory,\\
	23-28 June, 2014\\
	Columbia University New York, NY}
\begin{document}

\section{Introduction}
\label{sec:Intro}
The understanding of strong interactions is of fundamental importance to clarify the properties of elementary particles. 
Quarks form nucleons bound by the strong interaction and are hence considered to be the fundamental
constituents of matter.
A theory describing properties of the strong interaction is called Quantum ChromoDynamics (QCD).
Numerical simulations of QCD defined on the lattice have been used as a quantitative tool to
clarify non-perturbative properties of quarks and gluons. 
Lattice QCD simulations are based on Markov Chain Monte Carlo methods and have led to  
many important results such as the baryon spectrum \cite{Alexandrou:2014sha} and provided
essential insight into the
properties of the strong interactions.
Nevertheless, it is clear that conventional Lattice QCD simulations at finite chemical potential 
have only a limited range of applicability since 
the complex probability distribution at finite chemical potential spoils the Monte Carlo sampling. 
Therefore, it is one of the grand tasks to establish other numerical tools for investigations at
a large chemical potential. 
So far, several approaches have been tried among which are the complex Langevin equation, 
Lefschetz thimbles etc. 
Here, we focus on yet another alternative, the Hamiltonian approach combined with 
tensor networks techniques.

Tensor network states (TNS) can be a way to tackle the difficulties of conventional Monte Carlo methods.
TNS are efficient entanglement-based parameterizations of quantum many body states,
with a number of free parameters that typically grows polynomially with the system size, 
in contrast to the exponential growth of the dimension of the full Hilbert space.
In particular, Matrix Product States (MPS) are one-dimensional TNS,
which lie at the basis of the density matrix renormalization group (DMRG)
method~\cite{Schollwock2005,PhysRevLett.91.147902,PhysRevLett.93.227205}.
Interestingly, it can be proven that the ground states of local gapped Hamiltonians in 1D, and 
also the thermal equilibrium states can be well-approximated by MPS of moderate bond dimension~\cite{Hastings2007&PhysRevB.73.085115}.
From a more applied perspective, the MPS-based numerical algorithms allow us to 
systematically explore the \emph{physically relevant corner} of the Hilbert space.             
These methods are now well established and have been successfully applied to many  
condensed matter problems.

One of the earliest studies with a Hamiltonian approach was done with a strong coupling expansion
in~\cite{Banks:1975gq}.
Nowadays, this attempt has various branches:
TNS techniques~\cite{Banuls:2013jaa,Banuls:2013zva}, 
extension of the  strong coupling expansion~\cite{Cichy:2012rw},
numerical analysis with the DMRG approach~\cite{Byrnes:2002nv}, 
real-time evolution~\cite{Kuhn:2014rha, Buyens:2013yza},
and quantum link models~\cite{Rico:2013qya&Banerjee:2012xg&2013dda}. 
Another approach which works directly with the path integral is the  
tensor renormalization group~\cite{Shimizu:2014uva&Liu:2013nsa}.

This paper is organized as follows. We introduce the Schwinger model 
for $N_{\rm f}=1$, which is investigated as a testbed in the next section. 
The methods applied to the model in this study are explained in Sec.~\ref{sec:mthd}. 
In Sec.~\ref{sec:rslt}, we show some preliminary results.  
Finally, we conclude in Sec.~\ref{sec:cncl}.

\section{The Schwinger model for $N_{\rm f}=1$}
\label{sec:Schwgr}
The Schwinger model is 1+1 dimensional Quantum ElectroDynamics (QED). 
It has been investigated for a couple of decades,
since it has common features with QCD: a confinement property and chiral symmetry breaking via an anomaly. 
From the Lagrangian of the model, one can obtain the corresponding Hamiltonian. 
Using the Jordan-Wigner transformation, the Hamiltonian in a dimensionless form (${\tilde H}$)
can be written in spin language as~\cite{Banks:1975gq}
\begin{eqnarray}
\label{Hamil}
  {\tilde H}
  =\frac{2}{g^2a} H
  &=& x \displaystyle \sum_{n=0}^{N-2}
  \left[ \sigma_n^+ \sigma_{n+1}^- 
  + \sigma_n^- \sigma_{n+1}^+ \right]
  +\frac{\mu}{2} \sum_{n=0}^{N-1} 
  \Big[ 1+ (-1)^n \sigma_n^z \Big]
  + \sum_{n=0}^{N-2} \left[ L(n) \right] ^2\\
  &\equiv& {\tilde H}_{hop} + {\tilde H}_m + {\tilde H}_g,
\end{eqnarray}
where $x=1/g^2a^2$, $\mu= 2m/g^2a$, $m$ denotes the fermion mass, $a$ the lattice spacing, $g$ the coupling and $N$
the number of sites.
The gauge field, $L(n)$, can be treated in terms of the spin content by
integrating it out using Gauss' law: $L(n+1) - L(n) = \frac{1}{2} \left[ (-1)^{n+1} + \sigma_{n+1}^z \right]$.
We take $L(0)=0$.

We are interested in the chiral condensate of the Schwinger model at finite temperature. 
The temperature dependence of the chiral condensate was analytically found to be~\cite{Sachs:1991en}: 
\begin{eqnarray}
   \left\langle {\bar \psi}\psi \right\rangle 
   &=& \frac{m_{\gamma}}{2\pi} e^{\gamma} e^{2I(m_{\gamma}/T)} 
   = \left\{ \begin{array}{cl}  
                       \frac{m_{\gamma}}{2\pi} e^{\gamma}  & {\rm for} \hspace{1cm}  T\rightarrow 0  \\
                       2T e^{-\pi T/m_{\gamma}}                   & {\rm for} \hspace{1cm}  T\rightarrow \infty 
          \end{array}  \right.
   \label{eq:analytic}
\end{eqnarray}
where $I(x) = \int_0^{\infty} \frac{dt}{1-e^{x\cosh(t)}}$, $m_{\gamma}=g/\sqrt{\pi}$
and $\gamma = 0.57721\dots$ is Euler's constant. 
In Fig.~\ref{fig:dlt&D}, the analytic formula of the chiral condensate is shown as a black, solid curve. 

\section{Methods}
\label{sec:mthd}
\subsection{Matrix Product States}
In this paper, we employ the one dimensional Matrix Product States (MPS) ansatz. 
For a chain of $N$ sites, each one holding a physical degree of freedom with dimension $d$,
 an MPS is a state of the form
\begin{eqnarray}
   \left| \psi \right\rangle 
   &\approx& \displaystyle \sum_{i_0, i_1, \dots} {\rm Tr} \left[{M{[0]}}^{i_0} {M{[1]}}^{i_1} \dots {M{[N-1]}}^{i_{N-1}} \right] \left| i_0 i_1 \dots i_{N-1} \right\rangle
   \label{eq:MPS}
\end{eqnarray}
where $i_k=1,\dots,d$ is the physical index at site $k$ and $M[k]$ is a tensor defined at site $k$. 
Each tensor has three indices: one physical, with dimension $d$, and two virtual, of dimension
$D$, which is called the bond dimension.
The trace in Eq.~(\ref{eq:MPS}) refers to these virtual indices. 
The number of parameters in an MPS state is thus $N d D^2$.

The construction can be illustrated with a simple example for a state of two spin-$\frac{1}{2}$ particles.
We consider the following linear combination of two states.
\begin{eqnarray}
  \frac{1}{\sqrt{2}} \left( \left| \uparrow \downarrow \right\rangle + \left| \downarrow \uparrow \right\rangle \right) 
  &=& \displaystyle \sum_{i_0, i_1=\uparrow, \downarrow} {\rm Tr} \left[{M{[0]}}^{i_0} {M{[1]}}^{i_1} \right] \left| i_0 i_1 \right\rangle.
  \label{eq:ex_state}
\end{eqnarray}
The tensors $M{[0]}$ and $M{[1]}$ can be described by the following four $2\times 2$ matrices: 
\begin{eqnarray}
{M{[0]}}^{i_0= \uparrow}
= \left( \begin{array}{cc}
0 & \frac{1}{\sqrt{2}} \\
0 & 0 \end{array} \right), \hspace{1mm}
{M{[0]}}^{i_0= \downarrow}
= \left( \begin{array}{cc}
0 & 0 \\
\frac{1}{\sqrt{2}} & 0 \end{array} \right),  \hspace{1mm}
{M{[1]}}^{i_1= \uparrow}
= \left( \begin{array}{cc}
0 & 1 \\
0 & 0 \end{array} \right),  \hspace{1mm}
{M{[1]}}^{i_1= \downarrow}
= \left( \begin{array}{cc}
0 & 0 \\
1 & 0 \end{array} \right).  \hspace{1mm}
\label{eq:Ms}
\end{eqnarray} 
The MPS description is not unique, but it is subject to the so-called gauge freedom \cite{Perez2007mpsrepresentation}. 
The value of an individual coefficient for each basis element can be obtained 
as the trace of a product of $M$ tensors,
e.g. ${\rm Tr} \left[M[0]^{i_0=\uparrow} M[1]^{i_1=\uparrow} \right]=0$ in Eq.~(\ref{eq:ex_state}).  
Notice that this is an exact description of state~(\ref{eq:ex_state}) as an MPS.
In general, writing a given state as an MPS
can require a bond dimension up to $d^{\lfloor N/2\rfloor}$, 
i.e. scaling exponentially as the size of the Hilbert space. 
Nevertheless, it is often possible to find MPS of
small bond dimension that provide sufficiently good approximations to
the interesting states.
Theoretical results support the existence of such approximations
for ground and thermal states
under concrete conditions on the Hamiltonian~\cite{Hastings2007&PhysRevB.73.085115}.

In practice, the bond dimension can be treated as one parameter of the simulation, 
which obviously affects the computational effort
and controls the precision of the results.


\subsection{Global optimization}
A fundamental step in the algorithms is the transformation of an MPS under the action of a certain
operator,
${\mathcal O}$, e.g. an operator for real/imaginary time evolution. Starting from an initial MPS,
 $\left| \psi \right\rangle_{\rm init}$, the best MPS approximation, $\left| \psi \right\rangle_{\rm MPS}$, to the 
 product ${\mathcal O} \left| \psi \right\rangle_{\rm init}$
 can be found by minimizing the distance
 \begin{eqnarray}
   \Delta 
   &\equiv& \left\| {\mathcal O} \left| \psi \right\rangle_{\rm init} - \left| \psi \right\rangle_{\rm MPS} \right\|^2.
\end{eqnarray}
In our case, we consider ${\mathcal O}$ operators with an Matrix Product Operator (MPO) description 
(see~\cite{1367-2630-12-2-025012} for further details), 
and the minimization is performed by varying one tensor at a time, and 
sweeping back and forth over the chain until convergence.
This is called global optimization, to be distinguished from the TEBD \cite{PhysRevLett.91.147902} approach
in which only a few tensors of the chain are changed.

\subsection{Thermal calculation}
The thermal expectation value of the chiral condensate operator is given by
\begin{eqnarray}
   \Sigma 
   &\equiv& \frac{\left\langle {\bar \psi}\psi  \right\rangle}{g}
   = \frac{{\rm Tr} \left[{\hat \Sigma} \hspace{1mm} \rho(\beta) \right]}{ {\rm Tr} \left[ \rho (\beta) \right]}
\end{eqnarray}
where $\rho (\beta)$ is the thermal equilibrium density operator at inverse temperature $\beta=1/T$.
In spin language, the chiral condensate operator is expressed as 
${\hat \Sigma} = \frac{\sqrt{x}}{N} \sum_n (-1)^{n} \left[ \frac{1+\sigma_n^z}{2} \right]$.

The MPO approximation to the thermal state, $\rho(\beta)\propto e^{-\beta H}$,
can be computed using imaginary time evolution \cite{VerstraeteGarcia-RipollCirac2004}
acting on the identity matrix, which corresponds (up to normalization) to the exact thermal state at $\beta=0$,
and is an MPO of bond dimension 1. 
The exponential operator is applied using a
Suzuki-Trotter decomposition in which every imaginary time step, $\delta$, is
approximated by 
\begin{eqnarray}
   \exp \left(-\delta {\tilde H} \right) 
   &\approx& \exp\left(-\frac{\delta}{2} {\tilde H}_g\right) \exp\left(-\frac{\delta}{2} {\tilde H}_e\right) \exp\left(-\delta {\tilde H}_o\right) \exp\left(-\frac{\delta}{2} {\tilde H}_e\right) \exp\left(-\frac{\delta}{2} {\tilde H}_g\right)
   \label{eq:expH}
\end{eqnarray}
where ${\tilde H}_e$ (${\tilde H}_o$) is the part of the hopping term, ${\tilde H}_{hop}$, that contains terms starting on an even
(odd) site.
We consider the massless case, so that  the mass term ${\tilde H}_m$ is dropped.
Whereas the exponentials of ${\tilde H}_{e(o)}$ can be written exactly as MPO of small bond dimension, 
for the gauge part, which contains long range terms,  $\exp(-\delta {\tilde H}_g/2)$ is approximated by its Taylor expansion $1 - \frac{\delta}{2} {\tilde H}_g$. 

In order to find the best  approximated MPO for the thermal state at each inverse temperature, 
the initial MPO is vectorized, i.e. cast into an MPS form, $|\rho\rangle$ and the action of the evolution
operator (or any of its pieces as described above),
$U \rho U^{\dagger}$, is then mapped to $U\otimes U^{*} |\rho\rangle$.
Then, the MPS global optimization described above can be used iteratively 
to find the action of the desired number of evolution steps.
The resulting MPO will in general not be positive,
as the truncation procedure does not guarantee positivity. 
This is nevertheless easily achieved by simulating $\rho(\beta/2)$ 
and then constructing $\rho(\beta) = \rho(\beta/2) \rho^{\dagger}(\beta/2)$.

\section{Results}
\label{sec:rslt}
\begin{figure}[t] 
   \begin{minipage}{7cm}
   \includegraphics[width=7cm,clip]{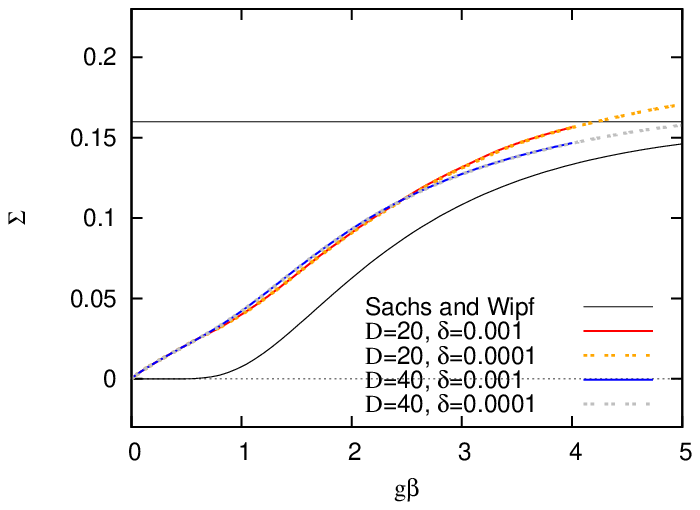} 
   \caption{Step size $\delta$ dependence of the chiral condensate at $D=20$ (red, orange), $40$ (blue, grey lines) for $x=16$: the black curve shows the analytic formula in~\cite{Sachs:1991en}, the black horizontal line is the low temperature limit of the formula.}
   \label{fig:dlt&D}
   \end{minipage}
   \hspace{5mm}
   \begin{minipage}{7cm}
   \centering
   \includegraphics[width=7cm,clip]{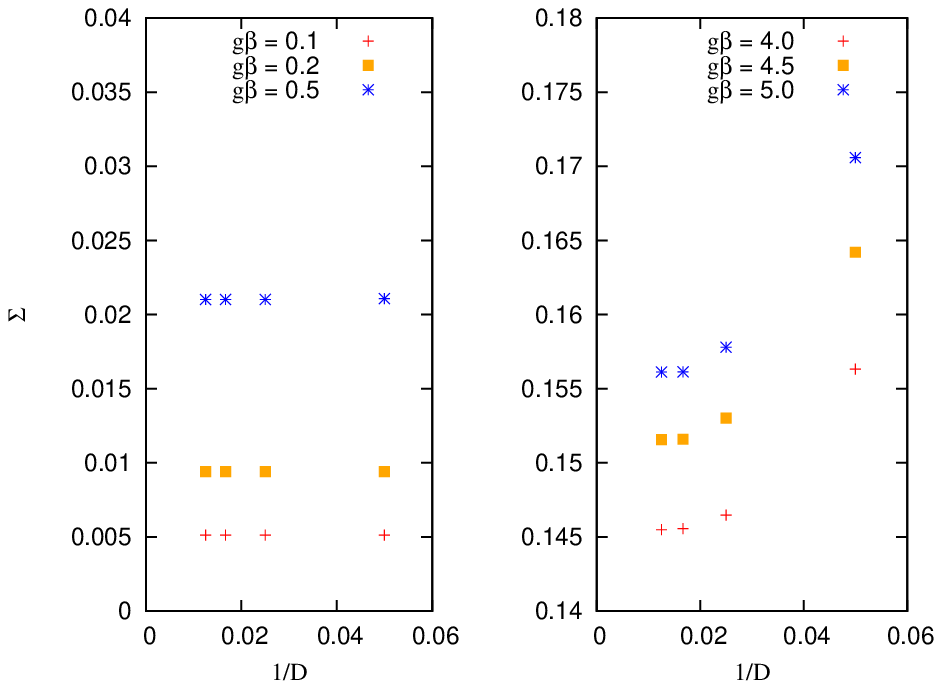} 
   \caption{Bond dimension $D$ dependence of the condensate in the high temperature region $0.1 \le g\beta \le 0.5$ (left) and the low temperature $4.0 \le g\beta \le 5.0$ (right).}
   \label{fig:D}
   \end{minipage}
   \hspace{5mm}
\end{figure}
\begin{figure}
   \begin{minipage}{7cm}
   \includegraphics[width=7cm]{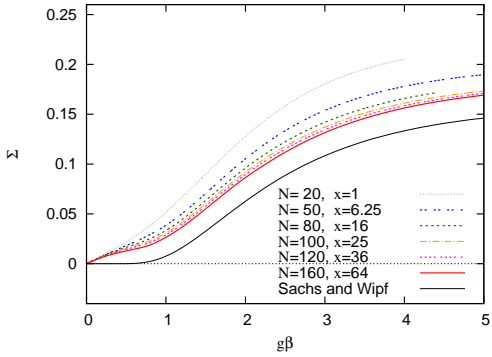} 
   \caption{Lattice spacing dependence of the condensate with fixed $N/\sqrt{x}=20$.} 
   \label{fig:fixed_N/sqrtx}
   \end{minipage}
   \hspace{0.8cm}
   \begin{minipage}{7cm}
   \vspace{-1.0cm}
     \tblcaption{Simulation setup for the continuum extrapolation: $D=80$ and $\delta=1\times 10^{-6}$ to $5 \times10^{-5}$.}
     \vspace{2mm}
     \label{tab:setup}
      \centering
      \begin{tabular}{cc} \hline
      $x$  &  $N$  \\ \hline
      25   &   80-140  \\
      36   &   80-160  \\
      49   &   80-180  \\
      64   &   80-220  \\
      81   &   120-220   \\
    121   &   180-240\\ \hline
      \end{tabular}
   \end{minipage}
\end{figure}
Using the techniques described above, we have computed 
the MPS approximation to the thermal state of the Schwinger model 
and evaluated the expectation value of the chiral condensate
for several system sizes and lattice constants.
Here, the temperature dependence is expressed as a function of $g\beta$ 
translated from dimensionless inverse temperature ${\tilde \beta}(\equiv \frac{g^2a}{2}\beta) = n\delta$ through the equation $g\beta=2\sqrt{x} {\tilde \beta}$.
Ideally, one should choose an infinitesimal value of the step size $\delta$
and sufficiently large value of the bond dimension $D$. 
The finite value of $\delta$ and $D$ causes systematic errors,
that needs to be controlled.
First of all, to estimate those errors, we perform a computation of the chiral condensate in the case of $N=20$ and $x=16$.
Fig.~\ref{fig:dlt&D} shows the step size $\delta$ dependence of the condensate for $D=20$ and $D=40$. 
It can be seen that the step size dependence is rather small compared to dependence on the bond dimension $D$. 
Additionally, in Fig.~\ref{fig:dlt&D}, one can see that the $D$ dependence appears 
more strongly 
in the low temperature region $g\beta \gtrsim 1.0$.
We check the convergence in the bond dimension $D$ in Fig.~\ref{fig:D}. 
The $D$ dependence is suppressed in the high temperature region, 
e.g. $0.01 \lesssim 1/D \lesssim 0.05$ for $g\beta = 0.1, 0.2, 0.5$, 
while the dependence is clearly visible at the lower temperature. 
Nevertheless, for the parameters used here, 
we find also convergence for $D\gtrsim 40$. 
Therefore, we conclude that the systematic errors from a finite bond dimension 
and a finite step size are visible in the low temperature region, 
while they are less serious in the high temperature region.  
Let us note that we are presently performing a more systematic and comprehensive 
study of these systematic effects for various temperatures. 
In Fig.~\ref{fig:fixed_N/sqrtx} we show 
the condensate at various lattice spacings, i.e. $1/\sqrt{x}$ with $x=1/g^2a^2$. 
Here, we choose $N$ and $x$ satisfying $N/\sqrt{x} = 20$, which means a fixed physical size. 
Moreover, we take a bond dimension of $D=80$ and step sizes $\delta=2\times 10^{-5}$ to $10^{-3}$.  
In Fig.~\ref{fig:fixed_N/sqrtx}, we show the lattice spacing dependence of 
the condensate. Clearly, our numerical results approach the analytical curve of \cite{Sachs:1991en}
towards the continuum limit. 
Comparing Fig.~\ref{fig:dlt&D} and Fig.~\ref{fig:fixed_N/sqrtx}, 
one notes that the systematic errors from bond dimension/step size are small compared to the systematic errors 
of a non-zero lattice spacing and finite size effects, especially at high temperature.

\begin{figure}[t] 
   \vspace{-3mm}
   \begin{minipage}{7cm}
   \includegraphics[width=7cm]{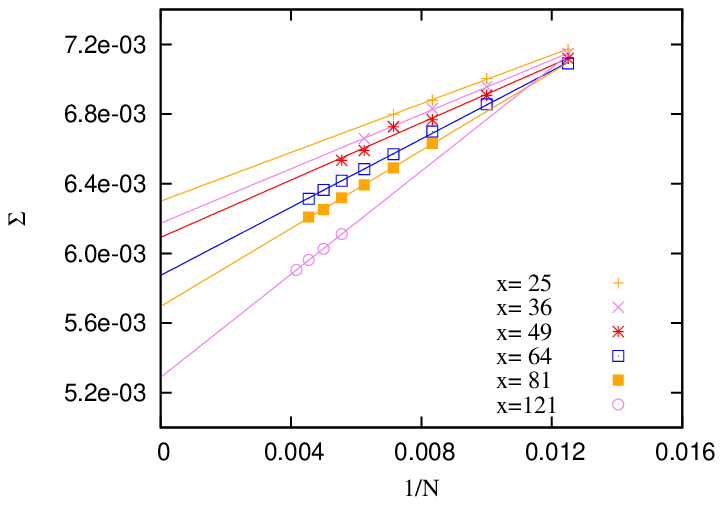} 
   \caption{Thermodynamic extrapolation with linear fit function in $1/N$ at different values of $x$ for $g\beta=0.2$.}
   \label{fig:InfVLimit}
   \end{minipage}
   \hspace{5mm}
   \begin{minipage}{7cm}
   \includegraphics[width=7cm]{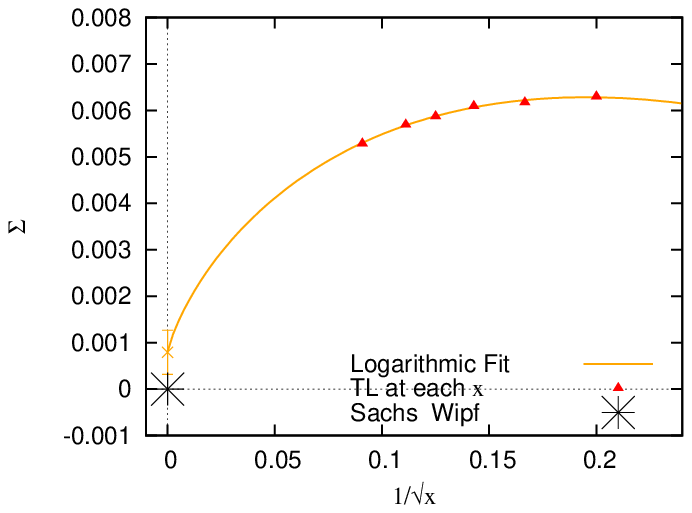} 
   \caption{Continuum limit extrapolation with logarithmic fit function for $g\beta=0.2$.}
   \label{fig:ContLimit}
   \end{minipage}
   \hspace{5mm}
   \vspace{-2.5mm}
\end{figure}
To control systematic errors appearing in our calculation, one has to perform four extrapolations:
vanishing step size limit $\delta \rightarrow 0$, large bond dimension limit, 
thermodynamic limit $N \rightarrow \infty$ and vanishing lattice spacing limit $1/\sqrt{x} \rightarrow 0$. 
As discussed above, for the parameters used here, we neglect the effects from 
$\delta$ and $D$ and only show the finite volume and lattice spacing effects.
Fig.~\ref{fig:InfVLimit} shows the infinite volume limit extrapolation 
using a linear fit in $1/N$ at $g\beta = 0.2$. 
The system sizes, $N$, used for each inverse coupling $x$ are summarized in
Tab.~\ref{tab:setup}. 
From these extrapolated values, we attempt to perform the continuum limit extrapolation at $g\beta = 0.2$
employing a theoretically motivated logarithmic fit \cite{Banuls:2013jaa,Banuls:2013zva}, 
as shown in Fig.~\ref{fig:ContLimit}.
We obtain a consistent result with the analytic calculation in~\cite{Sachs:1991en}.
We note that 
in Fig.~\ref{fig:InfVLimit} the thermodynamic limit extrapolation for $x=49$ 
does not show the expected linear behaviour. 
We interpret this observation by the fact that the finite values of $D$ and $\delta$ 
used here 
lead to a situation that we are simulating only 
an effective Hamiltonian which differs from the true model.
It is necessary to clarify those effects in order to investigate 
in particular the 
low temperature region and 
as mentioned above,  
therefore a more complete analysis is presently carried through.

\section{Conclusion}
\label{sec:cncl}
In this study, we have employed a tensor network (TN) approach to investigate the 
finite temperature behavior of the Schwinger model for the $N_{\rm f}=1$ flavour case.
In particular, we computed the temperature dependence of 
the chiral condensate. However, our method of the 
MPO approximation for the thermal state can be used more generally 
for the computation of
any MPO-like operator. 
From Figs.~\ref{fig:dlt&D} and \ref{fig:fixed_N/sqrtx}, we found that the systematic errors 
from a non-zero step size $\delta$ and from a finite bond dimension $D$ 
are rather small compared to those from a finite chain length $N$ and from 
a non-zero lattice spacing,
at least in the regime of high enough temperatures. 
Therefore, as a preliminary result, we neglected the former two systematic errors, 
but 
carried out the thermodynamic and the continuum limit extrapolation
in the high temperature region, shown in Figs.~\ref{fig:InfVLimit} and \ref{fig:ContLimit}. 
In this region, at $g\beta=0.2$, the result we obtain is consistent with the analytical 
prediction derived by 
Sachs and Wipf in~\cite{Sachs:1991en}.

\vspace*{3mm}
\noindent\textbf{Acknowledgments.} HS was supported by the Japan Society for the Promotion of Science for
Young Scientists. This work was partially funded by EU grand SIQS (FP7-ICT 2013-600645).


\begin{thebibliography}{99}
\bibitem{Alexandrou:2014sha}
  C.~Alexandrou, V.~Drach, K.~Jansen, C.~Kallidonis and G.~Koutsou,
  arXiv:1406.4310 [hep-lat].

\bibitem{Schollwock2005}
U.~Schollwock.
\newblock {\em Rev. Mod. Phys.} {\bf 77}, 259 (2005).

\bibitem{PhysRevLett.91.147902}
G.~Vidal.
\newblock{\em Phys. Rev. Lett.} {\bf 91}, 147902 (2003).

\bibitem{PhysRevLett.93.227205}
F.~Verstraete, D.~Porras and J.~I.~Cirac.
\newblock {\em Phys. Rev. Lett.} {\bf 93}, 227205 (2004).

\bibitem{Hastings2007&PhysRevB.73.085115}
M.~B.~Hastings.
\newblock {\em J. Stat. Mech. Theor. Exp.} {\bf 08}, 08024 (2007).
M.~B. Hastings.
\newblock {\em Phys. Rev.} {\bf B73}, 085115 (2006).

\bibitem{Banks:1975gq}
T.~Banks, L.~Susskind and J.~B.~Kogut.
\newblock {\em Phys. Rev.} {\bf D13}, 1043 (1976).

\bibitem{Banuls:2013jaa}
M.~C.~Ba{\~n}uls, K.~Cichy, K.~Jansen and J.~I.~Cirac.
\newblock {\em JHEP} {\bf 11}, 158 (2013).

\bibitem{Banuls:2013zva}
M.~C.~Ba{\~n}uls, K.~Cichy, J.~I.~Cirac, K.~Jansen and H.~Saito.
\newblock {\em PoS, Lattice2013} {\bf 332} (2013).

\bibitem{Cichy:2012rw}
K.~Cichy, A.~Kujawa-Cichy and M.~Szyniszewski.
\newblock {\em Comput. Phys. Commun.} {\bf 184}, 1666 (2013).
M.~Szyniszewski, K.~Cichy and A.~Kujawa-Cichy,
\newblock {\em PoS, Lattice2014} {\bf 314} (2014), arXiv:1410.7597 [hep-lat].

\bibitem{Byrnes:2002nv}
T.~Byrnes, P.~Sriganesh, R.~J.~Bursill and C.~J.~Hamer.
\newblock {\em Phys. Rev.} {\bf D66}, 013002 (2002).

\bibitem{Kuhn:2014rha}
S.~K\"{u}hn, J.~I.~Cirac and M.~C.~Ba{\~n}uls.
\newblock {\em Phys. Rev.} {\bf A90}, 042305 (2014).

\bibitem{Buyens:2013yza}
B.~Buyens, J.~Haegeman, K.~V.~Acoleyen, H.~Verschelde and F.~Verstraete.
\newblock {\em Phys. Rev. Lett.} {\bf 113}, 091601 (2013).

\bibitem{Rico:2013qya&Banerjee:2012xg&2013dda}
E.~Rico, T.~Pichler, M.~Dalmonte, P.~Zoller and S.~Montangero.
\newblock {\em Phys. Rev. Lett.} {\bf 112}, 201601 (2014).
D.~Banerjee, M.~Bogli, M.~Dalmonte, E.~Rico, P.~Stebler, et~al, 
\newblock {\em Phys. Rev. Lett.} {\bf 110}, 25303 (2013).

\bibitem{Shimizu:2014uva&Liu:2013nsa}
Y.~Shimizu and Y.~Kuramashi.
\newblock {\em Phys. Rev.} {\bf D90}, 014508 (2014).
Y.~Liu, Y.~Meurice, M.~P.~Qin, J.~Unmuth-Yockey, T.~Xiang, et~al.
\newblock {\em Phys. Rev.} {\bf D88}, 056005 (2013).

\bibitem{Sachs:1991en}
I.~Sachs and A.~Wipf.
\newblock {\em Helv. Phys. Acta} {\bf 65}, 652 (1992).

\bibitem{Perez2007mpsrepresentation}
D.~P\'erez-Garc{\'\i}a, M.~M.~Wolf, F.~Verstraete and J.~I.~Cirac.
\newblock {\em Quantum Inf. Comput.} {\bf 7}, 401 (2007).

\bibitem{1367-2630-12-2-025012}
B~Pirvu, V~Murg, J~I Cirac and F~Verstraete.
\newblock {\em New Journal of Physics}, {\bf 12}, 2, 025012 (2010).

\bibitem{VerstraeteGarcia-RipollCirac2004}
F.~Verstraete, J.~J.~Garc\'ia-Ripoll and J.~I.~Cirac.
\newblock {\em Phys. Rev. Lett.} {\bf 93}, 207204 (2004).

\end{thebibliography}



%
%
%
\end{document}